\documentclass{article}
\usepackage{spconf,amsmath,graphicx,hyperref}
\usepackage{enumitem}
\usepackage{amssymb}
\usepackage{booktabs}
\usepackage{algorithm}
\usepackage{algpseudocode}
\usepackage{bbm}
\usepackage{threeparttable}

\title{Asymmetric Classifier-Free Guidance for Target-Speaker ASR 
}
\name{Yiwen Guan, Jacob Whitehill}
\address{Worcester Polytechnic Institute, Massachusetts, USA }
\begin{document}
\ninept
\maketitle
\begin{abstract}
Target-speaker automatic speech recognition (TS-ASR) must identify and transcribe a desired speaker under varying overlap and noise conditions. 
These changes alter the acoustic evidence for the target speaker in the speech mixture, motivating inference-time calibration of speaker conditioning. 
We introduce asymmetric classifier-free guidance (CFG) for TS-ASR using Whisper: the speaker-conditioned branch predicts the target transcript, while the speaker-unconditioned branch predicts serialized multi-speaker transcripts. 
CFG adjusts the contribution of speaker conditioning during decoding through a single guidance scale. 
We select a global guidance scale on target-domain development data and train a lightweight encoder-based predictor to adjust it for each utterance, keeping the recognition model fixed.  
Under domain shifts, our full system achieves relative word error rate (WER) reductions of up to 21.8\% over the condition-only baseline, and 5.6\% over standard conditional decoding of the same CFG-trained model.  
Oracle analysis shows that substantially larger WER reductions are possible through utterance-level scale selection and identifies how beneficial adjustments vary with domain shifts. 

\end{abstract}
%


\section{Introduction}
\label{sec:intro}
Target-speaker automatic speech recognition (TS-ASR) transcribes a specific speaker from a mixture of multiple speakers, typically using an enrollment recording to identify the target. 
It is useful in applications such as voice assistants and meeting transcription, where a system needs to recognize one speaker in the presence of competing speech or ambient noise. 
Large-scale pretrained attention-based encoder-decoder models including Whisper \cite{radford2023robust} and OWSM v4 \cite{peng25c_interspeech} provide strong foundations for target-speaker recognition. 
By introducing a conditioning mechanism into Whisper, it is possible to harness its speech recognition capabilities for TS-ASR. 

Recent Whisper-based TS-ASR methods introduce speaker information through prompts or additional conditioning modules \cite{meng24c_interspeech, ma2024extending, guo2024sq}.
Alternatively, self-speaker adaptation derives this information from speech-activity-masked mixture features \cite{wang25y_interspeech}. 
While these methods are useful, a model trained on one set of mixtures may encounter different overlap patterns or noise levels at inference time. 
The conditioning learned during training may then need adjustment to maintain high accuracy. 
We thus explore whether a trained TS-ASR model can adapt to  domain shifts by adjusting the influence of speaker conditioning during decoding. 

\begin{figure*}[ht]
    \centering
    \includegraphics[width=1.0\linewidth]{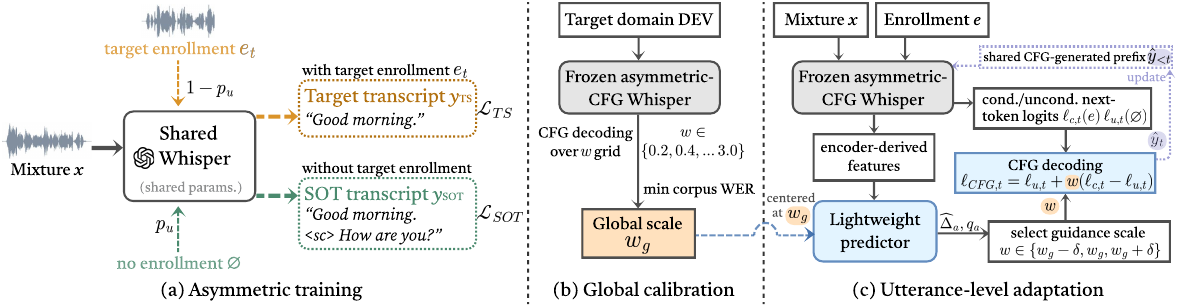}
    \vspace{-12pt}
    \caption{\textbf{Overview of the proposed framework.} 
    (a) A shared Whisper model is trained for TS-ASR with target enrollment and multi-speaker ASR without enrollment. 
    (b) A global guidance scale $w_g$ is calibrated on target-domain development data (DEV) with the ASR model frozen. 
    (c) A lightweight predictor performs utterance-level guidance adaptation around $w_g$, and the selected scale is applied to CFG decoding. 
    }
    \label{fig:framework}
\vspace{-8pt}
\end{figure*}

In this work, we introduce asymmetric classifier-free guidance (CFG) \cite{ho2021classifierfree} for Whisper-based TS-ASR. 
CFG is an inference-time procedure that combines predictions made with and without a conditioning signal through a guidance scale. 
In our model, the speaker-conditioned branch learns the target transcript, while the speaker-unconditioned branch learns serialized transcripts of all speakers, enabling both tasks within one shared backbone and giving a defined recognition task when the target identity is unavailable. 
At each decoding step, the difference between their predicted logits (see ``CFG decoding'' box in Figure \ref{fig:framework}) favors tokens preferred by the target-speaker branch relative to the multi-speaker branch. Scaling this difference provides a way to adjust that preference when acoustic conditions change; $w=1$ recovers ordinary speaker-conditioned decoding. 
To adapt to a new domain, we keep the trained backbone fixed and use a labeled target-domain development set to select a global guidance scale and train a lightweight predictor for utterance-level adjustment. 
To our knowledge, this is the first study to use CFG for inference-time calibration of TS-ASR under domain shift. 

Our research contributions are: 
    (1) We enable a single Whisper-based model to perform both target- and multi-speaker recognition by using asymmetric CFG training. 
    (2) We develop inference-time adaptation through development-set-selected global CFG calibration and a predictor for utterance-level adjustment. Global calibration improves test WER under domain shifts, and the predictor provides additional improvements. 
    (3) We analyze how domain shifts affect which guidance scales improve recognition. Oracle selections reveal substantial headroom beyond global calibration.


\section{Related Work}
\label{sec:relate}

Whisper-based TS-ASR systems use prompt tuning or speaker-querying modules to introduce target-speaker information  \cite{ma2024extending, guo2024sq}. 
Joint target- and multi-speaker recognition has also been achieved by combining Sidecar separation \cite{meng2023sidecar} with target-stream identification \cite{meng24c_interspeech}. 
In the related task of target speaker extraction (TSE), a Whisper-based generative TSE model trains TS-ASR as an auxiliary task and improves  performance of both tasks \cite{ma2025enhancing}. 
Other conditioning approaches use speaker activity through diarization-conditioned Whisper \cite{polok2026dicow}, or speaker-specific encoder kernels \cite{wang25y_interspeech}. 
Beyond Whisper-based systems, recent work explores joint activity detection \cite{maeda25_interspeech}, as well as LLM-based recognition with chain-of-thought reasoning and reinforcement learning \cite{zhang2025thinking, zhang2026xiaomi}. 
Optimizing speaker embeddings with ground-truth supervision reveals that the optimal embedding can depend on the mixture \cite{ashihara2024investigation}. CFG \cite{ho2021classifierfree} provides a feasible way to control conditioning at inference time. 
CFG has been used for domain adaptation during image generation \cite{buchheim2025controlling} and to strengthen acoustic conditioning in ASR \cite{huang2026ctc}. 
We use CFG to calibrate the contrast between target-speaker and multi-speaker predictions under domain shifts, which addresses the complementary question of how to calibrate speaker conditioning at inference time.


\section{Method}
\label{sec:method}

\subsection{Speaker-Conditioned Whisper} 
We built our TS-ASR model on Whisper, an attention-based encoder-decoder model pretrained on large-scale speech data and widely used for single-speaker tasks. 
To incorporate target-speaker information, we insert zero-initialized cross-attention adapters into selected Whisper encoder layers. Each adapter allows the mixture representation to attend to an enrollment representation of the target speaker, 
where the mixture representations serve as queries, while the enrollment representations serve as keys and values. 

Let $E(x,e)$ denote the resulting speaker-conditioned Whisper encoder, where $x$ is the speech mixture and $e$ is the speaker enrollment. 
The conditional and speaker-unconditioned encoder representations are  $H_c=E(x,e_t)$ and $H_u=E(x,\varnothing)$, 
where $e_t$ denotes the target-speaker enrollment and $\varnothing$ means each cross-attention adapter receives a zero vector in place of the enrollment sequence. 
The speaker-conditioned and unconditioned branches share all model parameters and are referred to hereafter as the conditional and unconditional branches. Here, ``unconditional" refers only to the absence of speaker enrollment; both branches remain conditioned on the speech mixture and generated prefixes. 
We also attach a linear CTC head \cite{10.1145/1143844.1143891} to the final encoder representation to provide frame-level alignment supervision. 
Both branches use a hybrid CTC/attention loss, as described in the following section. 

\subsection{Asymmetric CFG Training}

Classifier-free guidance (CFG) was originally developed for diffusion models to control the influence of a conditioning signal by combining conditional and unconditional predictions in image generation. It has been successfully applied to speech recognition \cite{huang2026ctc}, and also to autoregressive (AR) generation \cite{ICLR2026_d649dd58, ICLR2026_172be8b0}. 
We apply CFG to Whisper to adjust the influence of target-speaker conditioning during decoding without training auxiliary speaker identifiers. 

During training, the conditional branch $c$ is trained on the TS-ASR task with a hybrid loss: 
\begin{equation}
\label{eq:loss}
 \mathcal{L}_\text{TS}=
    \alpha\mathcal{L}_{\mathrm{CTC}}(H_\text{c}, y_\text{TS}^{\mathrm{CTC}})
    +(1-\alpha)\mathcal{L}_{\mathrm{CE}}(H_\text{c}, y_\text{TS}^{\mathrm{Attn}}),
\end{equation}
where $\alpha$ is set to 0.5 throughout this work. 
With probability $p_u$, the speaker enrollment is randomly removed, thus becoming the unconditional branch. 

Unlike conventional CFG training, which uses the same prediction target with and without conditioning, our unconditional branch is trained on a different objective: the serialized transcripts of both speakers with a special speaker-change token $\langle\mathrm{sc}\rangle$, aka serialized output training (SOT) \cite{kanda20b_interspeech}. 
Following standard permutation-invariant training \cite{yu2017permutation}, the lower loss of the two speaker orders is selected, denoted as $L_{\mathrm{SOT}}$. Thus, the proposed asymmetric CFG training becomes: 
\begin{equation}
    \mathcal{L}_{\mathrm{asym}}=\mathbb{E}_{b\sim\mathrm{Bernoulli}(p_u)}
\left[
(1-b)\mathcal{L}_{\mathrm{TS}}
+
b\mathcal{L}_{\mathrm{SOT}}
\right],
\label{eq}
\end{equation}
where $b\sim\mathrm{Bernoulli}(p_u)$ indicates whether the unconditional branch is selected for the current training sample. 

\subsection{Inference-time CFG Calibration and Adaptation}
\label{subsec:inference}

After asymmetric training, the same model produces target-speaker predictions when conditioned on speaker enrollment and serialized multi-speaker predictions when the enrollment is dropped. 
At inference time, we combine the predictions using CFG at every AR decoding step $t$. 
Both branches share the same generated prefix $\hat{y}_{<t}$ but attend to different encoder representations, $H_c$ and $H_u$. Let $\ell_{c,t}$ and $\ell_{u,t}$ denote the corresponding next-token logits, the guided next-token distribution is:
\begin{equation}
 p_w(y_t \mid \hat y_{<t},x,e_t)=
 \operatorname{softmax}
 \!\left[\ell_{u,t}+w\bigl(\ell_{c,t}-\ell_{u,t}\bigr)\right], 
 \label{eq:cfg}
\end{equation}
where $w$ is the guidance scale. Here $w=0$ gives unconditional multi-speaker decoding, $w=1$ recovers standard target-speaker decoding, and $w>1$ amplifies the difference between two predictions. 
This combination is applied at every decoding step, similar to context-aware AR decoding \cite{shi2024trusting}, while the guidance scale remains fixed within an utterance. 

The appropriate guidance strength can change when the test domain differs from the training domain. 
We therefore first perform global calibration by selecting the CFG scale $w_g$ that minimizes corpus WER over a predefined grid on target-domain development data. $w_g$ is fixed for test decoding, without updating the ASR model. 

Since the best $w$ may also vary across utterances, we further perform utterance-level adaptation with an action set
$\mathcal{A}=\{w_g-\delta,w_g,w_g+\delta\}$, where $\delta$ is the action spacing (step size).  
For each development example $i$ and a non-center action $a\in\mathcal{A}\setminus\{w_g\}$, we define: 
\begin{equation}
    \Delta_{i,a}=r_i(a)-r_i(w_g),
    \qquad
    b_{i,a}=\mathbbm{1}[\Delta_{i,a}<0],
    \label{eq:adaptive}
\end{equation} 
where $r_i(w)$ is the utterance WER obtained with guidance scale $w$, and $b_{i,a}$ indicates whether action $a$ improves WER over the global scale $w_g$. 
A lightweight predictor takes encoder-derived features 
and produces two outputs for each non-center action: the predicted WER change $\widehat{\Delta}_{i,a}$ and the benefit probability $q_{i,a}\in(0,1)$. 
We train $\widehat\Delta_{i,a}$ toward ${\Delta}_{i,a}$ with smooth-$L_1$ loss, and train $q_{i,a}$ toward $b_{i,a}$ using binary cross-entropy. 
After each training epoch, we evaluate pairs of thresholds $(\tau,m)$ on the validation subset. Each pair determines which guidance scale is selected for each validation utterance, and we compute the resulting corpus WER. 
The predictor adjusts the global scale only when both the confidence and predicted WER reduction exceed the thresholds. 
The final decoding procedure is detailed in Algorithm~\ref{alg:adaptive_cfg}. 
All predictor parameters, thresholds, and guidance scales remain fixed for evaluation.

\begin{algorithm}[t]
\caption{Global-centered adaptive CFG decoding}
\label{alg:adaptive_cfg}
\begin{algorithmic}[1]
\Require Mixture $x$, enrollment $e$, frozen ASR model $M$
\Require Predictor $f_\theta$, action set $\mathcal{A}$
\Require DEV-selected center $w_g$ and thresholds $(\tau,m)$
\State $z \gets \operatorname{EncoderFeatures}_{M}(x,e)$
\State $\{(\widehat{\Delta}_a,q_a)\}_{a\in
       \mathcal{A}\setminus\{w_g\}} \gets f_\theta(z)$
\State $a^* \gets \arg\min_{a\in
       \mathcal{A}\setminus\{w_g\}}\widehat{\Delta}_a$
\State $w \gets w_g$
\If{$q_{a^*}\ge\tau$ \textbf{and} $\widehat{\Delta}_{a^*}\le-m$}
    \State $w \gets a^*$
\EndIf
\State \Return $\operatorname{DecodeCFG}_M(x,e;w)$
\end{algorithmic}
\vspace{-3pt}
\end{algorithm}


\section{Experimental Setup}
\label{sec:experiment}

\subsection{Datasets and Evaluation Metrics} 
Our experiments are conducted on Libri2Mix \cite{cosentino2020librimix}, using the noise-free \texttt{mix\_clean} setting. The mixtures contain two speakers from LibriSpeech \cite{7178964}, with both recordings starting at the beginning of the mixture. Training uses the LibriSpeech \texttt{train-clean-100} subset, matching the training-data setting in \cite{meng24c_interspeech, ma2024extending}. 

To evaluate adaptation under controlled domain shifts, we construct development and test mixtures following the LibriSpeechMix setup \cite{kanda20b_interspeech}, referred to as LSM-controlled. 
The development and test sets contain 5,406 and 5,240 examples, respectively. 
We vary the overlap ratio across \{30\%, 50\%, 70\%\} (denoted OV30/50/70 for simplicity). Here we define the overlap ratio as the overlap of the two source recordings divided by the duration of the shortest recording. 
At OV50, we also add WHAM! noise \cite{wichern19_interspeech} at signal-to-noise ratios (SNR) of 10 and 0 dB. Global and adaptive CFG are evaluated on the three clean overlap settings, and OV50 at 10 and 0 dB. 

Following \cite{meng24c_interspeech}, we use a randomly cropped 3-sec LibriSpeech recording as speaker enrollment for each speaker. 
We report target-speaker word error rate (WER) for TS-ASR, and concatenated minimum-permutation WER (cpWER) \cite{watanabe20b_chime} for multi-speaker ASR. 
All  WERs are computed at the corpus level.

\subsection{Model and Calibration Settings} 
We use the pretrained Whisper-small model as the backbone throughout the experiments, with enrollment cross-attention adapters inserted after Whisper encoder layers 3, 6, and 9. 
During training, the enrollment is dropped with a probability of $p_u=0.1$, and the hybrid-loss weight is $\alpha=0.5$. 
All ASR models are trained for 60 epochs on one NVIDIA L40S GPU with a batch size of 8, using the AdamW optimizer \cite{loshchilov2018decoupled}. The learning rates are 1e-5 for the pretrained Whisper parameters and 1e-4 for newly added parameters. 

For each target-domain setting, we select the global guidance scale $w_g$ by minimizing development-set corpus WER over a grid from 0.2 to 3.0 in increments of 0.2. 
The ASR backbone is fixed throughout the calibration and predictor training stage. 
The adaptive CFG predictor is a multilayer perceptron with two 256-unit hidden layers and a dropout 0.1, trained for 10 epochs with a learning rate of 1e-3. 
The predictor uses encoder features from the speech mixture and speaker enrollment, which summarize the conditional and unconditional representations and their differences. 
Predictor training and validation use disjoint development set groups: 85\% of the development examples for training the predictor and the remaining 15\% for validation.  
The predictor checkpoint and decision thresholds are selected jointly by the corpus WER on the validation subset (as described in Sec.~\ref{subsec:inference}) and stay fixed for evaluation. 
All comparisons among backbone models are over 4 runs, while the global and adaptive CFG experiments use a fixed backbone and predictor seed.

\begin{table}[t]
\centering
\caption{
Libri2Mix-clean test WER (\%) with Whisper-small. For condition-only and CFG systems, we report mean ${\pm}$ sample standard deviation over four runs. Our CFG systems use $w=1$ for TS-ASR; asymmetric CFG uses $w=0$ for multi-speaker ASR. Our best run achieves 15.66\% TS-WER. Each `+' denotes a separate fine-tuning setup starting from pretrained Whisper. }
\vspace{3pt}
\label{tab:backbone}
\small
\setlength{\tabcolsep}{4pt}
\resizebox{0.9\columnwidth}{!}{%
\begin{threeparttable}
    \begin{tabular}{lcc}
        \toprule
        System & TS-WER & cpWER\\
        \midrule
        \multicolumn{3}{l}{\textit{Published Whisper-small TS-ASR}}\\
        TS-Whisper LoRA~\cite{ma2024extending} & 19.06 & --\\
        Whisper-SS-TTI-limited~\cite{meng24c_interspeech} & 15.75 & --\\
        Whisper-SS-TTI~\cite{meng24c_interspeech} \tnote{$\dagger$} & 11.81\tnote{$\dagger$} & 9.39\tnote{$\dagger$} \\
       
        \midrule
        Whisper (pretrained) & 69.26 & -- \\
        + SOT-only & -- & 14.10\\
        + Condition-only& 18.04${\pm}$1.85 & --\\
        + TS-CFG-sym, $p_u$=0.1& 17.61${\pm}$1.98 & -- \\
        + TS-CFG-asym, $p_u$=0.3 & 18.65${\pm}$1.80 & 21.97${\pm}$8.56 \\
        + TS-CFG-asym, $p_u$=0.1 (ours) & 17.59${\pm}$1.96 & 23.51${\pm}$4.97 \\
        \bottomrule
    \end{tabular}
    
    \begin{tablenotes}
      \small
      \item[$\dagger$] Uses additional training data beyond our limited-data setting; not directly comparable. 
    \end{tablenotes}
\vspace{-8pt}
\end{threeparttable}
}
\end{table}


\section{Results}
\label{sec:results}

\subsection{Target- and Multi-Speaker Recognition}

Table~\ref{tab:backbone} compares our trained backbone with published Whisper-small-based systems and our training variants on Libri2Mix-clean, before inference-time calibration. 
We evaluate our system with $w=1$ for TS-ASR, corresponding to standard conditional decoding, and $w=0$ for multi-speaker recognition. 
Across four independent training runs with different random seeds, our model achieves $17.59 \pm 1.96\%$ TS-WER, compared with $18.04 \pm 1.85\%$ for condition-only training and $17.61 \pm 1.98\%$ for symmetric CFG training, where $\pm$ denotes the sample standard deviation.  
Our best run achieves $15.66\%$ TS-WER, comparable to the 15.75\% of Whisper-small-SS-TTI-limited with the same backbone and training-data setting in \cite{meng24c_interspeech}. 

Our model also supports multi-speaker recognition, achieving $23.51 \pm 4.97\%$ cpWER. Although this remains higher than our SOT-only baseline's $14.10\%$, the model supports both recognition tasks while retaining TS-ASR performance close to the symmetric training baseline. 
Increasing the unconditional training probability from $p_u$=0.1 to 0.3 improves mean cpWER from 23.51\% to 21.97\%, but also increases mean TS-WER. 
This indicates a trade-off between target-speaker and multi-speaker recognition with respect to the choice of $p_u$. 
The benefit of asymmetric training lies in the joint capability and the basis it provides for inference-time calibration.

\begin{table}[t]
\centering
\caption{Test WER (\%) on LSM-controlled. Development-set-calibrated global scale ($w_g$) is selected and frozen for test set evaluation. Utterance-level adaptive CFG selects among three scales around the same global center ($\{w_g-\delta$,$w_g$,$w_g+\delta\}$). 
}
\vspace{3pt}
\label{tab:adaptive}
\small
\setlength{\tabcolsep}{3pt}
\begin{tabular}{l|c|cc|cc}
\toprule
 \multicolumn{1}{r|}{Backbone} & Cond. only& \multicolumn{4}{c}{TS-CFG-asym}\\
Setting &  & $w$=1 & $w_g$ & \multicolumn{2}{c}{Adaptive} \\
OV / SNR &  &  &  & $\delta$=0.4 & $\delta$=0.6 \\
\midrule
OV30          &  41.28&36.22 & 34.78 & 34.34 & \textbf{34.18} \\
OV50          &  32.23&31.47 & 30.94 & 30.69 & \textbf{30.65} \\
OV70          &  30.14&25.79 & 25.69 & 25.50 & \textbf{25.34} \\
OV50 / 10 dB   &  43.87&35.75 & 34.74 & 34.39 & \textbf{34.30} \\
OV50 / 0 dB   & 87.77 & 74.16 & 71.28 &   \textbf{70.50}& 71.06\\
\bottomrule
\end{tabular}
\vspace{-6pt}
\end{table}

\begin{table}[t]
\centering
\caption{Post-hoc oracle results on LSM-controlled-OV50. Both global and utterance-level oracle scales are selected using test references over the full guidance grid. 
Parentheses show the corresponding $w_g$ or $w_g^\text{oracle}$ for each setting. 
Switch rate is the percentage of examples for which some scale in the grid has better WER than $w$=1. 
}
\vspace{3pt}
\label{tab:oracle}
\small
\setlength{\tabcolsep}{3pt}
\resizebox{0.9\columnwidth}{!}{%
\begin{tabular}{l|cc|c|c|c}
\toprule
  &  && \multicolumn{3}{c}{Test set full-grid oracle results}\\
SNR &\shortstack{$w$=1\\WER} &\shortstack{$w_g$\\WER ($w$)}&  \shortstack{Test $w_g^\text{oracle}$ \\ WER ($w$)}
&\shortstack{Utt. oracle \\ WER}& Switch rate\\
\midrule
+$\infty$           &31.47 &30.94 (1.4)&  30.78 (1.6)&21.60& 17.2\%\\
10 dB   &35.75 &34.74 (1.8)&  34.59 (1.6)&25.46& 24.2\%\\
0 dB   &74.16 &71.28 (2.4)&  70.90 (2.2)&59.64& 38.4\%\\
\bottomrule
\end{tabular}
}
\vspace{-3pt}
\end{table}

\subsection{Global and Utterance-Level CFG Adaptation}

Table~\ref{tab:adaptive} evaluates global and utterance-level CFG adaptation under overlap (OV30/50/70) and noise shifts (OV50 with SNR=10 and 0 dB). 
As mentioned in Sec.\ref{subsec:inference}, for each setting, the development-set-selected global scale ($w_g$) is frozen for test set evaluation, and considered as the global center. Utterance-level adaptive CFG further selects among three actions based on the global center.  

Even without inference-time guidance adjustment ($w$=1), asymmetric TS-CFG outperforms the condition-only baseline across all five settings, indicating improved robustness with asymmetric CFG training. 
With the backbone fixed, applying $w_g$ further reduces test WER by 0.1--2.88\%. 
The largest WER reduction occurs at OV50/0\,dB, from 74.16\% down to 71.28\%. 
Among the clean overlap settings, OV30 benefits most, with WER decreasing from 36.22\% to 34.78\%. 
We speculate that this is because OV30 entails the largest domain shift among all controlled overlap settings from the left-aligned training mixtures, whose overlap ratio is 100\% under our definition. The same holds for SNR=0\,dB since the training set is clean. 
These results suggest that adjusting the condition guidance strength on target domains rather than maintaining a fixed scale can improve recognition without updating the ASR model. 

Further, utterance-level adaptation improves upon the global calibration in all five settings. 
With action spacing $\delta$=0.4, WER decreases by 0.19--0.78\%, and with $\delta$=0.6, WER decreases by 0.22--0.60\%. 
This also outperforms the best fixed scale selected using test reference (oracle) over the global search grid (70.9\% in Table~\ref{tab:oracle}) by 0.4\%. 
Together, global calibration and utterance-level adaptation reduce WER from 36.22\% to 34.18\% on OV30 with $\delta$=0.6, reducing over standard conditional decoding ($w$=1) by 5.6\% relatively. 
Thus, a lightweight predictor can use utterance-specific information to further improve on a shared global scale, while the effect of action spacing $\delta$ varies across conditions.

\subsection{Oracle Headroom Analysis}
Table~\ref{tab:oracle} uses test references to select either one fixed guidance scale $w_g^\text{oracle}$ for the entire test set or a separate scale for each utterance over the full grid (from 0.2 to 3.0 in increments of 0.2). 
At OV50, the development-set-selected global scales $w_g$ yield WERs only 0.15--0.38\% above the test global $w_g^\text{oracle}$ results. 
Both development and test set selections favor stronger guidance at 0 dB, supporting the use of development data to calibrate guidance under noise shifts. 
Utterance-level oracle selection provides a further 9.13--11.26\% reduction over the best fixed scale $w_g^\text{oracle}$, revealing substantial headroom beyond global calibration, which stems from the variation in the guidance strength preferred by individual utterances. 

As noise increases, the oracle switch rate rises from 17.2\% in clean mixtures to 24.2\% at 10 dB and 38.4\% at 0 dB. The best fixed test-set scale $w_g^\text{oracle}$ also increases from 1.6 to 2.2 at 0 dB. 
Together, these observations suggest that noise changes both the preferred global guidance scale and how often utterances benefit from adjusting the guidance strength from $w=1$. 
The lightweight CFG predictor already improves on global calibration, while the oracle results show substantial room for further improvement.

\begin{figure}[t]
    \centering
    \includegraphics[width=1.0\columnwidth]{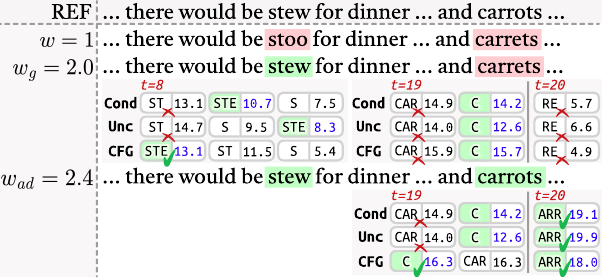}
    \vspace{-18pt}
    \caption{An example of CFG global calibration and utterance-level adaptation improving upon standard conditional decoding ($w$=1). 
    The wrong words are marked red, and the correct words/tokens are marked green. The tables under global $w_g$=2.0 and adaptive $w_{ad}$=2.4 display the top logits predicted by the conditional (Cond), unconditional (Unc), and the computed $\mathrm{CFG}=\text{Unc}+w\cdot(\text{Cond}-\text{Unc})$ at a specific decoding time step $t$. }
    \label{fig:example}
\vspace{-8pt}
\end{figure}


\section{Conclusion}
\label{sec:conclusion}

In this work, we introduce asymmetric CFG for Whisper-based TS-ASR, supporting target-speaker and multi-speaker recognition within a shared model while maintaining competitive TS-ASR performance. 
Even before inference-time calibration, our model outperforms condition-only baselines under domain shifts, indicating improved robustness. 
With the recognition model fixed, selecting a global guidance scale on target-domain development data improves test WER under changes in overlap and noise. 
A lightweight CFG predictor provides further gains through utterance-level guidance adjustment. 
Our analysis shows how beneficial adjustments vary with acoustic conditions, while oracle selection reveals substantial headroom beyond global calibration. 
These results demonstrate the effectiveness of CFG for inference-time adaptation in TS-ASR.
Our approach on inference-time control is complementary to advances in speaker conditioning and backbone design, and could extend to other TS-ASR architectures. 
Future work will explore better utterance-level predictors to close the oracle gap, as well as token-level guidance for finer control during decoding.



\textbf{Compliance with Ethical Standards.} 
This research study was conducted retrospectively using human subject data made available in open access by the LibriSpeech corpus (CC BY 4.0). Ethical approval was not required as confirmed by the license attached with the open access data.



\bibliographystyle{IEEEbib}
\bibliography{strings,refs}

\end{document}